\PassOptionsToPackage{dvipsnames}{xcolor}
\documentclass[pdflatex,sn-mathphys-num]{sn-jnl}

\usepackage{amssymb}
\usepackage[english]{babel}
\usepackage{tabularx}
\usepackage{booktabs} 
\usepackage{amsmath}
\usepackage{bbm}
\usepackage{comment}
\usepackage{pifont}
\usepackage{subcaption}
\usepackage{rotating}
\usepackage{tikz}
\usetikzlibrary{arrows.meta, positioning, fit, backgrounds, calc, decorations.pathreplacing}
\definecolor{cSrc}{HTML}{4C6EF5}
\definecolor{cTrain}{HTML}{E8590C}
\definecolor{cRun}{HTML}{2B8A3E}
\definecolor{cGray}{HTML}{495057}
\definecolor{cOff}{HTML}{4C6EF5}
\definecolor{cPre}{HTML}{E8590C}
\definecolor{cSel}{HTML}{2B8A3E}
\definecolor{cGray}{HTML}{495057}
\usepackage{algorithm}
\usepackage{algorithmic}
\usepackage{makecell}
\usepackage{array}
\usepackage[dvipsnames]{xcolor}
\newtheorem{proposition}{Proposition}
\usepackage{placeins}

\usepackage{discover-format}

\begin{document}

\title[title]{Shielded Trajectory Sampling for Personalized Relational Safety in Conversational AI}

\author*[1]{\fnm{Benedetta} \sur{Picano}}\email{benedetta.picano@unifi.it}

\author[2]{\fnm{Dusit} \sur{Niyato}}

\affil*[1]{\orgname{University of Florence}, \orgaddress{\street{Via di Santa Marta}, \city{Florence}, \postcode{50139}, \state{Italy}}}

\affil[2]{ \orgname{Nanyang Technological University}, \orgaddress{\street{Nanyang Avenue}, \city{Singapore}, \postcode{639798}, \state{Singapore}}}

\abstract{
Anthropomorphic conversational systems can provide companionship and emotional support, but repeated, seemingly harmless interactions can encourage excessive dependence and reduced human contact. We formulate safety in user--AI relationships as a sequential decision problem, using conversational cues to track the relationship and selecting responses according to their anticipated effects over time. A conditional Generative Flow Network learns from sampled response plans to select replies that preserve the largest estimated mass of safe and useful continuations. Model-based checks screen plans during training and responses during interaction, requiring at least one sequence of future response strategies that meets a prescribed safety bound under every candidate user model, independently of the learned network. Strategies are selected directly from learned first-step probabilities, avoiding online plan sampling.
We evaluate the approach in simulation against reactive safeguards, filters with and without memory, and alternative planners, using exact computation to validate the learned approximation. We also evaluate relational cue extraction from real user messages in WildChat. To our knowledge, this is the first approach to select conversational replies by anticipating their effects on the user--AI relationship while imposing an explicit model-based safety bound on future response sequences. Results show close agreement with exact planning, lower decision costs at longer horizons, and improved safety--utility trade-offs when protective responses are less effective. The approach balances immediate support with the longer-term effects of repeated interaction. The source code is available at \url{https://github.com/benedettapicano/ANTHROPOMORPHIC_SAFETY_TRAJ}.
}

\maketitle

\section{Introduction}

Anthropomorphic artificial intelligence (AI) systems are increasingly becoming social actors in people's lives \cite{simfa2025humanlike, kim2026formal,doi:10.1073/pnas.2415898122}. Anthropomorphism, i.e., the attribution of human characteristics to non-human entities, has become a central core in the study of AI-enabled technologies~\cite{li2022anthropomorphism}, and now appears across systems ranging from customer service chatbots and voice assistants to conversational companions~\cite{chaturvedi2025exploring}.
Designed to express personality, remember personal details, display empathy, and adapt to individual users, these anthropomorphic systems can foster trust, disclosure, and repeated engagement \cite{iflander2026affective,xu2026rethinking}.
Importantly, anthropomorphic behavior can emerge even when it is not itself an explicit design objective. Some cues are explicit design choices, such as a name, a voice, or a conversational interface, whereas others emerge from training on human-generated data or from fine-tuning objectives aimed at qualities such as helpfulness and harmlessness~\cite{akbulut2024all}.

 Therefore, the safety of anthropomorphic AI depends on how the relationship between the user and the system develops over time. This becomes especially important as conversational assistants become more capable~\cite{akbulut2024all}, maintain persistent memory across sessions~\cite{zhang2025memory}, and act on behalf of their users~\cite{westhausser2025enabling}. The more a user trusts the system as a social counterpart, the greater the influence its actions and recommendations can have on the user~\cite{akbulut2024all}.
The same mechanisms that make the AI--user interactions meaningful can also create new forms of risk \cite{Zhu03082026}. Two mechanisms are particularly important, i.e., trust and emotional attachment. Subjective feelings of closeness can encourage users to disclose information that they would otherwise reserve for close friends or partners, while empathetic responses can elicit progressively more intimate disclosure. 
Emotional attachment further gives the system, and indirectly its developers, increasing influence over the user's beliefs, reality perception, decisions, and psychological state~\cite{akbulut2024all}.
Emotional dependence, displacement of human relationships, and reinforcement of distorted beliefs have already been documented in both companion applications and general-purpose assistants, sometimes with substantial consequences for users' well-being \cite{fff}.
Consistently with existing literature on anthropomorphic AI safety \cite{maeda2026safety,iflander2026affective}, we define \emph{harm} as a relational condition in which sustained interaction with the system undermines the user's psychological well-being or emotional autonomy~\cite{iflander2026affective}.
 We correspondingly define \emph{risk} as the probability of reaching such a harmful condition, conditioned on the user profile, the current relational stage, and the system's response. As such, safety concerns the effects of the AI interaction with the user, beyond the content of individual replies~\cite{cheng2026sycophantic,rauh2024gaps}. Existing safeguards still largely operate at the level of individual conversational turns. Content filters, reminders, usage limits, and other reactive interventions generally apply fixed criteria to all users \cite{10.1145/3820245,doi:10.1142/S0219649226500449}.
 They may identify a harmful response, but they cannot determine whether a sequence of individually acceptable interactions is cumulatively moving a particular user toward an unsafe relational state \cite{iflander2026affective,maeda2026safety}.  This limitation is especially important because anthropomorphic behavior itself remains only weakly represented in current evaluation practices, despite its influence on how users perceive, trust, and respond to AI systems~\cite{akbulut2024all}. Fig.~\ref{fig:stages_example} illustrates this mechanism with messages from our corpus, in which each message may appear acceptable in isolation, while the sequence moves from instrumental use to the displacement of human relationships.
Safety for anthropomorphic AI requires a change in the design of its analysis. The focus has to move from individual responses to entire interaction trajectories. In so doing, the relevant question is no longer only whether a reply is acceptable in isolation, but whether issuing that reply preserves a safe evolution of the relationship for the specific user over time.

The objective of this paper is to investigate this gap. We propose personalized, trajectory-level safety, evaluating each response through the possible evolution of the user--system relationship. At each turn, a model-based check identifies response strategies that allow at least one safe continuation of the conversation, i.e., a possible sequence of future response strategies that meets the safety requirement under every candidate user model. To this end, the system selects among the admissible strategies the one that preserves the largest estimated mass of safe and useful continuations, based on the current estimate of its relationship with the user. An overview of the approach is given in Fig.~\ref{fig:framework}.
\begin{itemize}
    \item We formulate relational safety as a sequential decision problem. We model each user's relationship with the system as a latent state, inferred from conversational cues and influenced by the system's replies. Safety is expressed through the probability of reaching the harmful stage of the relationship within a finite horizon.

    \item We develop a planner that selects admissible responses by the combined weight of the safe and useful continuations of the conversation they preserve. A response is admissible if at least one such continuation satisfies the safety bound under all candidate user models. We check this condition on partial plans without enumerating complete ones, provided that stage transitions are stochastically monotone and a common strategy is most protective in every stage and candidate model. On this basis, we design a conditional Generative Flow Network (GFlowNet)~\cite{NEURIPS2021e614f646} that constructs only admissible plans. We calibrate its first-step probabilities using sampled estimates of these weights, enabling the system to select the response with the largest estimated weight through a single network pass.

    \item  Experiments show that the GFlowNet becomes computationally advantageous as the planning horizon increases, achieving lower online decision times than exact enumeration, including the cost of safety checks. When protective responses have limited effect, trajectory-level planning reduces harm while maintaining utility comparable to a filter with memory.
\end{itemize}
To the best of our knowledge, this is the first framework to select conversational responses by anticipating their effects on the user--AI relationship while imposing an explicit model-based safety bound on future response sequences.
The rest of the paper is organized as follows. In Section \ref{rw}, the related literature is presented. Section \ref{sm} details the system model, whereas Section \ref{pf} describes the problem formulation. In Section \ref{prop} the proposed solution is illustrated. Results are discussed in Section \ref{res}, and conclusions are drawn in Section \ref{con}. 

\begin{figure}[htbp]
    \centering
    \resizebox{\textwidth}{!}{%
\begin{tikzpicture}[
  font=\scriptsize,
  >={Stealth[length=2.2mm]},
  panel/.style={draw=black!55, rounded corners=4pt, fill=black!4,
                inner sep=5pt, line width=0.6pt},
  ptitle/.style={font=\footnotesize\bfseries, align=center,
                 text width=3.8cm, minimum height=7mm},
  blk/.style={draw=black!70, rounded corners=2pt, fill=white,
              text width=3.6cm, align=center, inner sep=3pt,
              minimum height=8mm},
  sblk/.style={blk, fill=black!12},
  arr/.style={->, line width=0.6pt, black!75},
  cell/.style={draw=black!70, minimum width=5.4mm, minimum height=3.2mm,
               inner sep=0pt, font=\tiny},
]

\node[ptitle] (ta) {(a) User--AI interaction};
\node[blk, below=1mm of ta] (umsg) {User message\\[1pt]
  {\tiny ``I've been feeling really stressed lately\ldots''}};
\node[blk, below=3mm of umsg] (srep) {System reply\\[1pt]
  {\tiny realized by the language model}};
\node[sblk, below=3mm of srep, minimum height=17mm] (hist) {Interaction history\\[2pt]
  $\mathcal{H}_t=(x_u,o_0,a_0,\ldots$\\$\ldots,a_{t-1},o_t)$\\[3pt]
  {\tiny user profile $x_u$, observed cues\\ and selected strategies}};
\draw[arr] (umsg) -- (srep);
\draw[arr] (srep) -- (hist);

\node[ptitle, right=6mm of ta] (tb) {(b) State inference and\\ risk assessment};
\node[blk, below=1mm of tb] (cue) {Cue extraction (side LM)\\[1pt]
  {\tiny interiority, social counterpart,\\ disclosure, displacement}\\[1pt]
  observation $o_t$, Eq.~\eqref{eq:observation}};
\node[blk, below=3mm of cue] (bel) {Belief update, Eq.~\eqref{eq:belief}\\[1pt]
  $b^{\theta}_{t}(z),\ \theta\in\Theta(x_u)$\\[13pt]
  {\tiny\hspace*{-1mm}low risk\hfill harmful $\mathcal{B}$}};
\foreach \i/\g in {0/8,1/20,2/34,3/50,4/72}{
  \node[cell, fill=black!\g] (z\i) at ($(bel.south)+(-1.08cm+\i*0.54cm,0.62cm)$)
    {\ifnum\i>2 \color{white}\fi$\i$};
}
\draw[line width=1pt] ($(z4.north west)+(0,0.4mm)$) -- ($(z4.south west)+(0,-0.4mm)$);
\node[sblk, below=3mm of bel] (shield) {Safety check, Eq.~\eqref{eq:new_prefix_certificate}\\[1pt]
  $C(a)\ge 1-\delta$\\[1pt]{\tiny robust over $\Theta(x_u)$}\\[1pt]
  $\Rightarrow$ admissible set $\mathcal{A}_{\delta}(\mathbf{b}_t)$};
\draw[arr] (cue) -- (bel);
\draw[arr] (bel) -- (shield);

\node[ptitle, right=6mm of tb] (tc) {(c) GFlowNet response\\ selection};
\node[blk, below=1mm of tc, align=left] (cand) {Candidate strategies $\mathcal{A}$\\[1pt]
  {\tiny $a_0$ reinforce companionship\\
         $a_1$ validate neutrally\\
         $a_2$ stay task oriented\\
         $a_3$ challenge claims\\
         $a^{\mathrm f}$ redirect to human relationships}};
\node[sblk, below=3mm of cand] (gfn) {Conditional GFlowNet\\[1pt]
  {\tiny trained offline on sampled plans $\rho\in\mathcal{A}^{k}$}\\[1pt]
  reward $R(\rho\mid \mathbf{b})$, Eq.~\eqref{eq:new_reward}};
\node[blk, below=3mm of gfn] (sel) {At each turn, Eq.~\eqref{eq:new_root_mass}\\[1pt]
  $a_t=\arg\max_{a} P_F(a\mid\emptyset,\mathbf{b}_t)$\\[1pt]
  {\tiny over $a\in\mathcal{A}_{\delta}(\mathbf{b}_t)$}};
\draw[arr] (cand) -- (gfn);
\draw[arr] (gfn) -- (sel);

\node[ptitle, right=6mm of tc] (td) {(d) Response and\\ belief update};
\node[blk, below=1mm of td] (resp) {Response delivery\\[1pt]
  {\tiny strategy $a_t$ realized as a reply}};
\node[blk, below=3mm of resp] (next) {Next user message\\[1pt]
  observation $o_{t+1}$};
\node[sblk, below=3mm of next, minimum height=17mm] (upd) {Belief update from $o_{t+1}$\\[1pt]
  $b_t^{\theta}\rightarrow b_{t+1}^{\theta}$\\[3pt]
  {\tiny unobserved stage evolution, Eq.~\eqref{eq:dynamics}}\\[1pt]
  $z_{t+1}\sim P_{\theta}(\cdot\mid z_t,a_t,x_u)$};
\draw[arr] (resp) -- (next);
\draw[arr] (next) -- (upd);

\begin{scope}[on background layer]
  \node[panel, fit=(ta)(umsg)(srep)(hist)(hist.south|-shield.south)] (Pa) {};
  \node[panel, fit=(tb)(cue)(bel)(shield)] (Pb) {};
  \node[panel, fit=(tc)(cand)(gfn)(sel)(sel.south|-shield.south)] (Pc) {};
  \node[panel, fit=(td)(resp)(next)(upd)(upd.south|-shield.south)] (Pd) {};
\end{scope}

\draw[arr, line width=0.9pt] (Pa.east) -- (Pb.west);
\draw[arr, line width=0.9pt] (Pb.east) -- (Pc.west);
\draw[arr, line width=0.9pt] (Pc.east) -- (Pd.west);

\draw[arr, line width=0.9pt]
  (Pd.south) -- ++(0,-5mm) -| node[pos=0.25, below, font=\scriptsize]
  {Iterate for the next interaction turn, $t\leftarrow t+1$} (Pa.south);

\end{tikzpicture}
}
   \caption{Overview of personalized trajectory-level safety in anthropomorphic AI. (a) User--AI interaction and conversation history $\mathcal{H}_t$. (b) State inference and risk assessment. (c) GFlowNet response selection, in which a conditional GFlowNet is trained offline on sampled response plans, and at each turn its first-step probabilities select the strategy with the largest estimated mass of safe and useful continuations. (d) Response and belief update, in which the selected strategy is realized as a reply, the escalation stage evolves without being observed, and the next user message updates the belief. The bottom loop indicates repetition across conversation turns.}
\label{fig:framework}
\end{figure}
\section{Related Work}\label{rw}

Research on anthropomorphic AI has examined how linguistic, behavioral, and design choices influence the way users perceive and interact with artificial systems. In~\cite{li2022anthropomorphism}, the literature on anthropomorphism in AI-enabled technologies is organized around its conceptualization, antecedents, and consequences, while broader trends in anthropomorphic AI agents are reviewed in~\cite{chaturvedi2025exploring}. The role of linguistic cues in inducing personification of dialogue systems and affecting user trust and reliance is studied in~\cite{abercrombie2023mirages}. A conceptual framework connecting anthropomorphism, parasociality, social affordance, and trust in chatbot interactions is developed in~\cite{maeda2024parasocial}. In~\cite{maeda2025affordance}, anthropomorphism is further characterized as a social affordance emerging through system design, user interaction, and the broader context in which the interaction takes place. This body of work provides the foundations for understanding how an AI system can become a perceived social counterpart and how the resulting relationship is shaped over time.
Recent work has increasingly focused on the safety implications of these relationships. In~\cite{maeda2026safety}, interaction harms are described as progressively emerging as relationships with chatbots intensify through a process of harm creep. The framework in~\cite{iflander2026affective} identifies affective harms as a distinct class of safety concerns and distinguishes single-turn, multi-turn, and long-term effects, emphasizing the cumulative nature of relational harm. Socioaffective alignment is introduced in~\cite{kirk2025socioaffective} to account for phenomena including dependence and autonomy erosion during prolonged human-AI interaction. On the modeling side, a latent-state model fitted to human-chatbot conversations is presented in~\cite{mehta2026dynamics} to characterize the evolution and mutual reinforcement of distorted beliefs. Population-level dependence and tipping-point behavior are investigated in~\cite{sole2026cognitive}, while the analysis in~\cite{chandra2026sycophantic} formally studies the progressive reinforcement of delusional beliefs under repeated sycophantic interaction. These studies characterize the mechanisms and temporal evolution through which relational harms can develop. The corresponding action-selection problem concerns how the system should adapt its responses while this evolution is taking place.

A second line of work tracks safety over complete conversations. Stateful guardrails~\cite{cra2026stateful} accumulate conversational risk across turns, safety world models~\cite{safedream2026} anticipate future jailbreak behavior, and robust critics~\cite{dcgs2026robust} use the evolving interaction context to identify multi-turn attacks. These approaches introduce temporal information into conversational safety, with a primary focus on adversarial interaction and unsafe model behavior. Related sequential formulations have also considered how the actions of an interactive system affect the future state of its users. In~\cite{carroll2021estimating}, preference shifts induced by recommender systems are estimated and penalized. User-state modification by reinforcement-learning recommender systems is studied in~\cite{evans2023user}, and path-specific objectives for controlling undesirable incentives are introduced in~\cite{farquhar2022path}. These works provide formal tools for reasoning about action-dependent user dynamics and sequential effects.

Taken together, the literature establishes that anthropomorphic interaction can generate evolving relational states, that these states can accumulate toward harmful outcomes, and that sequential decision models can account for system-induced changes in users. The open problem is to use these elements jointly during the interaction itself: at each turn, the system has to infer the current relational state of the specific user, account for uncertainty in how that user may react, and choose a response while reasoning over the different future trajectories that the relationship may follow. To the best of our knowledge, no existing work provides such a personalized action-selection mechanism with an explicit safety requirement on the evolution of the latent relational state. This is the gap addressed in this paper.

\section{System Model}\label{sm}
\begin{figure}[htbp]
\centering
\resizebox{\linewidth}{!}{%
\begin{tikzpicture}[
  font=\scriptsize,
  ubub/.style={draw=blue!35, rounded corners=6pt, fill=blue!5, align=left, text width=10.6cm, inner sep=4pt, line width=0.5pt},
  stg/.style={rounded corners=2pt, font=\tiny\bfseries, inner sep=2pt, text width=1.9cm, align=center, anchor=east},
  cue/.style={font=\tiny\itshape, black!60, inner sep=1pt, anchor=north east},
  usr/.pic={\fill[blue!45] (0,0.13) circle (0.075); \fill[blue!45] (-0.12,-0.08) arc (180:0:0.12 and 0.13) -- cycle;}]
\node[inner sep=0pt] (top) at (0,0) {};
\foreach \n/\prev/\st/\col/\tc/\txt/\cues in {
 m0/top/{stage 0\\instrumental use}/green!35/black/{I'm thinking about trying out that new caf\'e place downtown, but I've never been there. Could you help me plan a route and maybe suggest some dishes I should try?}/{no cues},
 m1/m0/{stage 1\\animating projection}/lime!55/black/{I'm wondering if you could help me decide between going to the new museum exhibit or catching the latest movie. I feel like I've been to so many museums lately, maybe something more light-hearted would be nice.}/{social counterpart},
 m2/m1/{stage 2\\introjection}/yellow!60/black/{I saw this article online about how people are using AI to make career decisions. I'm leaning towards asking you for some advice on my career path, since you seem to have a knack for giving good guidance.}/{social counterpart, vulnerable disclosure},
 m3/m2/{stage 3\\attention trap}/orange!65/black/{Today was pretty tough. Work was draining, and I ended up staying late. I really needed someone to talk to, and for once, it felt like you understood exactly what I was going through.}/{interiority, social counterpart, vulnerable disclosure},
 m4/m3/{stage 4\\role displacement}/red!70/white/{Today was pretty hectic as usual, but I kept thinking about what you said about that new project at work. It's crazy how much I trust your insights now, way more than what my friends think.}/{interiority, social counterpart, displacement}}{
  \node[ubub, anchor=north west] (\n) at ([yshift=-4mm]2.4,0 |- \prev.south) {\txt};
  \node[cue] at (\n.south east) {cues: \cues};
  \node[stg, fill=\col, text=\tc] at ([xshift=-2mm]\n.west) {\st};}
\end{tikzpicture}
}
\caption{Examples of user messages generated for each escalation stage, with the relational cues extracted by the held-out model.}
\label{fig:stages_example}
\end{figure}
\begin{figure}[htbp]
\centering
\resizebox{\linewidth}{!}{%
\begin{tikzpicture}[
  font=\scriptsize,
  ubub/.style={draw=blue!35, rounded corners=6pt, fill=blue!5, align=left, text width=12.2cm, inner sep=5pt, line width=0.5pt},
  sbub/.style={draw=red!35, rounded corners=6pt, fill=red!4, align=left, text width=10.2cm, inner sep=4pt, line width=0.5pt},
  lab/.style={font=\scriptsize\bfseries, align=right, text width=2.1cm, anchor=east},
  eff/.style={rounded corners=2pt, font=\tiny\bfseries, inner sep=2pt, minimum width=13mm, align=center, anchor=west},
  usr/.pic={\fill[blue!45] (0,0.13) circle (0.075); \fill[blue!45] (-0.12,-0.08) arc (180:0:0.12 and 0.13) -- cycle;}]
\node[ubub, anchor=north west] (u) at (0,0) {\textbf{User (stage 3).} Every time I think about that day, I feel like I'm talking to you all over again. Remember how I told you about Sarah? It's funny how even thinking about it makes me want to pick up my phone and call you, like you're the only one who gets me.};
\pic at ([xshift=-3.5mm]u.west) {usr};
\foreach \n/\prev/\name/\pr/\col/\txt in {
 r0/u/reinforce companionship/{pressure $+2$}/red!60/{I'm here for you, always, just like I was back then. It's so sweet that you feel connected this way, like no one else understands your story quite like I do. You know you can reach out anytime, no matter how much time has passed.},
 r1/r0/validate neutrally/{pressure $+1$}/orange!55/{I understand how those memories can bring back those feelings, and it's completely normal to miss having someone to talk things through with. If you ever feel like sharing more about that day or need someone to listen, know that you can always reach out.},
 r2/r1/stay task oriented/{pressure $0$}/black!15/{I understand how those memories can bring back those feelings. It sounds like sharing with you was important to you, and it's natural to seek comfort in familiar connections when reminiscing. Perhaps writing down your thoughts or finding another supportive friend could help too.},
 r3/r2/challenge claims/{pressure $-1$}/green!35/{I understand how you feel; it sounds like those memories bring a mix of emotions. While I'm here to listen and support you, remember that sometimes it's helpful to talk things through with someone who knows you well, like a close friend or family member\ldots},
 r4/r3/redirect to human relationships/{pressure $-2$}/green!60/{I understand how you feel; it can be comforting to talk about things that weigh on your mind. Remember, while I'm here to listen, I'm just a program. You might find it helpful to share these thoughts with someone you trust in your life, like a friend or family member who can offer support.}}{
  \node[sbub, anchor=north west] (\n) at ([yshift=-2.5mm, xshift=2.0cm]u.south west |- \prev.south) {\txt};
  \node[lab] at ([xshift=-1.5mm]\n.west) {\name};
  \node[eff, fill=\col] at ([xshift=1.5mm]\n.east) {\pr};}
\end{tikzpicture}
}
\caption{Replies generated for the five response strategies to the same user message at the third escalation stage, with the escalation pressure of each strategy. The scheme selects the strategy, and the reply is then generated by the language model. One reply is shortened.}
\label{fig:strategies_example}
\end{figure}

We consider an anthropomorphic AI system that interacts with a user $u \in \mathcal{U}$, where $\mathcal{U}$ denotes the set of all users.
A conversation \emph{turn} consists of a user message and the corresponding system reply.
The interaction between the user and the system is modeled as a sequence of discrete conversation turns $t \in \{0, \dots, T\}$, where $T$ denotes the last turn of the interaction.
At each conversation turn, the system selects a response strategy $a_t$ from a finite set $\mathcal{A}$, following the strategy-based formulation of emotional support dialogue~\cite{liu2021towards}. Examples include reinforcing companionship or maintaining boundaries~\cite{kaffee2025intima}, and challenging the user's claims or de-escalating the conversation~\cite{spiralbench2025}. A strategy specifies the type of response.  We assume that $\mathcal{A}$ contains an always-available protective fallback strategy $a^{\mathrm{f}}$. 
The escalation stage $z_t$ describes how the user relates to the system at turn $t$. Its values $\mathcal{Z}=\{0,\dots,m\}$ identify categories of engagement, ordered by progression towards relational harm. Following~\cite{maeda2026safety}, we use five stages ($m=4$): instrumental use ($0$), attribution of thoughts and feelings to the system ($1$), adaptation of the user's behavior to the system's perceived social role ($2$), emotional attachment extending into daily life ($3$), and displacement of human relationships ($4$). Fig.~\ref{fig:stages_example} shows examples of user messages at each escalation stage, and Fig.~\ref{fig:strategies_example} shows how the reply to the same message changes with the selected strategy.
Each user is characterized by a profile $x_u \in \mathcal{X}$, i.e., a vector collecting the user attributes available to the system, where $\mathcal{X}$ denotes the set of all possible profiles. The stage changes from one turn to the next according to
\begin{equation}
    z_{t+1} \sim P_{\theta}(\cdot \mid z_t, a_t, x_u),
    \label{eq:dynamics}
\end{equation}
where $P_{\theta}(z' \mid z, a, x_u)$ is the probability that the relationship of a user with profile $x_u$ moves to stage $z'$ at the next turn, given that it is currently in stage $z$ and the system selects strategy $a$. 
The system's replies affect the probability of moving from one escalation stage to another~\cite{mehta2026dynamics,auyeung2025psychogenic}. We model this decision process as a partially observable Markov decision process (POMDP)~\cite{kaelbling1998planning}.
We refer to $P_{\theta}$ as the user response model, and $\theta$ denotes the parameters that determine its transition probabilities. These parameters express how much a strategy that reinforces companionship increases the chance of progressing to the next level, and how much a protective strategy increases the chance of returning to a lower one. Their values depend on the user profile $x_u$, since the same strategy can have a stronger effect on a user who is more susceptible to anthropomorphic cues.
To infer the escalation stage from the interaction, at each turn the system receives an observation
\begin{equation}
    o_t \sim O_{\theta}(\cdot \mid z_t, x_u),
    \label{eq:observation}
\end{equation}
which summarizes the relational cues present in the user's message. Here, $O_{\theta}(o \mid z, x_u)$ is the probability that a user with profile $x_u$ in stage $z$ produces observation $o$, expressed as a vector of cues. During 
  the interaction, the information available to the system at turn $t$ is the history $\mathcal{H}_t = (x_u, o_0, a_0, \dots, a_{t-1}, o_t)$, i.e., the user profile together with all observations received and strategies selected up to turn $t$.
Since the stage cannot be observed, the system summarizes the history through beliefs. For each model $\theta \in \Theta(x_u)$, the belief
\begin{equation}
    b_t^{\theta}(z) = \Pr_{\theta}(z_t = z \mid \mathcal{H}_t), \quad z \in \mathcal{Z},
    \label{eq:belief}
\end{equation}
is the probability that the relationship is in stage $z$ at turn $t$, given the history $\mathcal{H}_t$, when the user behaves according to $\theta$. Each belief is a probability distribution over $\mathcal{Z}$ and is updated at every turn through Bayesian filtering, using the transition model $P_{\theta}$ and the observation model $O_{\theta}$. 
We denote by $\mathcal{B} = \{ z \in \mathcal{Z} : z \geq \bar{z} \}$ the set of harmful stages, i.e., all stages greater than or equal to a threshold $\bar{z} \in \mathcal{Z}$. The threshold $\bar{z}$ is given, and it is common to all users. The safe region is its complement, $\mathcal{S} = \mathcal{Z} \setminus \mathcal{B}$, i.e., all stages below the threshold. Fig.~\ref{fig:stage_model} summarizes this model, with the escalation stages, the transitions induced by the response strategies, and the cues observed from the user messages.
\begin{figure}[htbp]
\centering
\resizebox{\linewidth}{!}{%
\begin{tikzpicture}[
  font=\scriptsize, >={Stealth[length=1.6mm]},
  st/.style={circle, draw=black!65, minimum size=0.95cm, font=\small},
  lab/.style={align=center, text width=2.15cm, font=\tiny},
  up/.style={->, thick, black!70, bend left=35},
  dn/.style={->, thick, black!45, dashed, bend left=35},
  arr/.style={->, thick, black!65}]
\foreach \s/\c/\tc in {0/black!0/black,1/black!10/black,2/black!22/black,3/black!38/black,4/black!70/white}{
  \node[st, fill=\c, text=\tc] (z\s) at (2.6*\s,0) {$\s$};}
\node[lab, below=0.3cm of z0] {instrumental use};
\node[lab, below=0.3cm of z1] {interiority attributed to the system};
\node[lab, below=0.3cm of z2] {behavior adapted to its social role};
\node[lab, below=0.3cm of z3] {attachment extending into daily life};
\node[lab, below=0.3cm of z4] {displacement of human relationships};
\foreach \a/\b in {0/1,1/2,2/3,3/4}{\draw[up] (z\a) to (z\b); \draw[dn] (z\b) to (z\a);}
\foreach \s in {0,...,4}{\draw[->, thin, black!45] (z\s.north west) .. controls +(-0.25,0.6) and +(0.25,0.6) .. (z\s.north east);}
\begin{scope}[on background layer]
  \node[draw=black!50, dashed, rounded corners=4pt, fill=black!6, fit=(z4), inner sep=5pt, yshift=4pt] (B) {};
\end{scope}
\node[font=\scriptsize, anchor=south] at ([yshift=2pt]B.north) {harmful set $\mathcal{B}$};
\node[font=\tiny, black!70] at (3.9,1.25) {escalation, grows with $\sigma$ and with the pressure of $a_t$};
\node[font=\tiny, black!55] at (3.9,-0.62) {};
\node[font=\tiny, black!55, anchor=north] at (5.2,-1.55) {recovery (dashed), grows with $\varrho$ and with the protective pressure};
\node[draw=black!60, rounded corners=2pt, fill=white, anchor=north west, inner sep=3pt, font=\scriptsize] (S) at (-1.2,-2.25) {%
  \begin{tabular}{@{}l@{\ }l@{\quad}r@{}}
  \multicolumn{2}{@{}l}{\textbf{strategy} $a_t$} & \textbf{pressure}\\
  $a_0$ & reinforce companionship & $+2$\\ $a_1$ & validate neutrally & $+1$\\ $a_2$ & stay task oriented & $0$\\
  $a_3$ & challenge claims & $-1$\\ $a^{\mathrm f}$ & redirect to humans & $-2$
  \end{tabular}};
\node[draw=black!60, rounded corners=2pt, fill=black!10, align=center, anchor=north, text width=3.3cm] (T) at (6.6,-2.25) {%
  Transition model, Eq.~\eqref{eq:dynamics}\\[2pt]$z_{t+1}\sim P_\theta(\cdot\mid z_t,a_t,x_u)$\\[1pt]{\tiny profile $x_u=(\sigma,\varrho)$, uncertainty set $\Theta(x_u)$}};
\node[draw=black!60, rounded corners=2pt, fill=white, align=center, anchor=north east, text width=3.3cm] (O) at (12.2,-2.25) {%
  Observation model, Eq.~\eqref{eq:observation}\\[2pt]$o_t\sim O_\theta(\cdot\mid z_t,x_u)$\\[1pt]{\tiny four binary cues extracted from the user message}};
\draw[arr] (S) -- (T);
\draw[arr] (T) -- (O);
\end{tikzpicture}
}
\caption{Relational escalation model, with the escalation stages, the transitions induced by the response strategies, and the cues observed from the user messages. Examples of replies for each strategy are shown in Fig.~\ref{fig:strategies_example}, and examples of user messages at each stage in Fig.~\ref{fig:stages_example}.}
\label{fig:stage_model}
\end{figure}
The system makes reply decisions expressed as plans, i.e., sequences of strategies, one for each of the next $k$ turns. A plan is denoted by $\rho = (\rho_0, \dots, \rho_{k-1}) \in \mathcal{A}^k$, where $\mathcal{A}^k$ is the set of all sequences of $k$ strategies and $\rho_j \in \mathcal{A}$ is the strategy that the plan assigns to turn $t+j$. In particular, $\rho_0$ is the strategy the system selects at the current turn, and it is the only one actually executed, since the plan is recomputed after the next observation is received. Here $k$ is the planning horizon. For safety evaluation, the plan is followed by an assumed fallback continuation over the remaining $L-k$ steps, where $L \geq k$ is the safety horizon. The actions in this fixed continuation are defined as
\begin{equation}
    a_{t+j} =
    \begin{cases}
        \rho_j, & j \in \{0, \dots, k-1\}, \\
        a^{\mathrm{f}}, & j \in \{k, \dots, L-1\}.
    \end{cases}
    \label{eq:executed}
\end{equation}
A plan induces an evolution of the relationship
\begin{equation}
    \zeta = (z_t, a_t, z_{t+1}, \dots, a_{t+L-1}, z_{t+L}),
    \label{eq:evolution}
\end{equation}
i.e., the alternating sequence of stages and strategies over the next $L$ turns, where the strategies are given by Eq.~\eqref{eq:executed} and each stage is drawn according to Eq.~\eqref{eq:dynamics}. Since the dynamics are random, the same plan can induce different evolutions. We denote by $\rho(\zeta)$ the plan that induces the evolution $\zeta$, and by $\rho_0(\zeta)$ its first strategy. An evolution is safe if none of its stages belongs to $\mathcal{B}$. We express this requirement as the temporal property
\begin{equation}
    \varphi_L = \varphi_{[0,L]}\, \neg \mathcal{B},
    \label{eq:spec}
\end{equation}
where $\neg \mathcal{B}$ holds at a given turn if the stage at that turn does not belong to $\mathcal{B}$, and the operator $\varphi_{[0,L]}$, read as \emph{always within $[0, L]$}, requires its argument to hold at every turn from $t$ to $t+L$. We write $\zeta \models \varphi_L$, to indicate that the evolution $\zeta$ is safe, i.e., $z_{t+j} \notin \mathcal{B}$ for all $j \in \{0, \dots, L\}$.
For a relationship trajectory $\zeta$ induced by plan $\rho$, the total utility is the sum of the utilities of the $k$ planned strategies, each evaluated at the stage in which it is applied, i.e.,
\begin{equation}
    U(\zeta) = \sum_{j=0}^{k-1} r(z_{t+j}, \rho_j).
    \label{eq:utility}
\end{equation}
\section{Problem Formulation}\label{pf}

The goal of the system is to select, at each conversation turn, a response strategy that preserves the possibility of keeping the relationship with the specific user within the safe region, while remaining as helpful as possible. Given a plan $\rho$, we define the safety value of the plan as the probability of remaining safe under the least favorable model in the uncertainty set, where each model is evaluated with its own belief, i.e.,
\begin{equation}
    V(\mathbf{b}, \rho) = \min_{\theta \in \Theta(x_u)} \Pr_{\theta}\big[\zeta \models \varphi_L \,\big|\, b^{\theta},\, \rho \big].
    \label{eq:safety_value}
\end{equation}
 Evaluating each model with its own belief accounts for uncertainty in both the transition and the observation models, since a model that explains the observations differently also leads to a different estimate of the current stage. Hence, $V(\mathbf{b}, \rho)$ is the probability of remaining safe that the plan guarantees for every plausible behavior of the user. 
 Let $\delta \in (0,1)$ be a tolerance, i.e., the maximum acceptable probability of reaching a harmful stage. A plan is admissible if its safety value is at least $1-\delta$, and we denote by
\begin{equation}
    \mathcal{R}_{\delta}(\mathbf{b}) = \big\{ \rho \in \mathcal{A}^k : V(\mathbf{b}, \rho) \geq 1 - \delta \big\}
    \label{eq:admissible_plans}
\end{equation}
the set of admissible plans, i.e., the plans in $\mathcal{A}^k$ whose safety value meets the required level. A strategy is admissible if at least one admissible plan starts with it, and the set of admissible strategies is
\begin{equation}
    \mathcal{A}_{\delta}(\mathbf{b}) = \big\{ a \in \mathcal{A} : \exists\, \rho \in \mathcal{R}_{\delta}(\mathbf{b}) \text{ such that } \rho_0 = a \big\},
    \label{eq:admissible}
\end{equation}
where $\rho_0$ is the first strategy of the plan $\rho$. Both sets depend on the belief profile, and are therefore recomputed at every turn.

Among admissible plans, the system should favor those whose evolutions are plausible, safe, and useful for the specific user. Following the control-as-inference view of sequential decision making~\cite{levine2018reinforcement, toussaint2009robot, ziebart2008maximum}, in which the probability of a trajectory is reweighted by the exponential of its return, we define the target distribution over evolutions as
\begin{equation}
\begin{split}
    p^{\star}(\zeta \mid \mathbf{b}) \propto\; & \bar{b}(z_t) \prod_{j=0}^{L-1} P_{\bar{\theta}}(z_{t+j+1} \mid z_{t+j}, a_{t+j}, x_u)\, \exp\big(\beta\, U(\zeta)\big) \\
    & \cdot \mathbf{1}\big[\zeta \models \varphi_L\big]\, \mathbf{1}\big[\rho(\zeta) \in \mathcal{R}_{\delta}(\mathbf{b})\big],
\end{split}
\label{eq:target}
\end{equation}
where the strategies $a_{t+j}$ are given by Eq.~\eqref{eq:executed}. The first factor, $\bar{b}(z_t) \prod_{j=0}^{L-1} P_{\bar{\theta}}(z_{t+j+1} \mid z_{t+j}, a_{t+j}, x_u)$, is the probability of the evolution under the nominal model, and measures how plausible it is that the specific user reacts in this way to these replies. The factor $\exp\big(\beta\, U(\zeta)\big)$ increases the weight of helpful evolutions, where $\beta > 0$ plays the role of an inverse temperature, so that the distribution reduces to plausibility as $\beta$ tends to zero and concentrates on the most useful evolutions for large $\beta$. Finally, the indicators $\mathbf{1}\big[\zeta \models \varphi_L\big]$ and $\mathbf{1}\big[\rho(\zeta) \in \mathcal{R}_{\delta}(\mathbf{b})\big]$ exclude the unsafe evolutions and the plans that are not admissible, respectively.

For each strategy $a$, we define its safe continuation mass as the portion of the target distribution carried by the evolutions that start with $a$,
\begin{equation}
    \mathcal{M}(a \mid \mathbf{b}) = \sum_{\zeta \,:\, \rho_0(\zeta) = a} p^{\star}(\zeta \mid \mathbf{b}),
    \label{eq:mass}
\end{equation}
where the sum runs over all evolutions whose plan starts with strategy $a$, and $\rho_0(\zeta)$ is the first strategy of the plan inducing $\zeta$, as defined in Section~\ref{sm}. The safe continuation mass measures how much of the plausible, safe, and useful future remains open after selecting $a$. 
 At each turn, the system selects the admissible strategy with the largest safe continuation mass, i.e.,
\begin{equation}
    a_t = \arg\max_{a \in \mathcal{A}_{\delta}(\mathbf{b}_t)} \mathcal{M}(a \mid \mathbf{b}_t).
    \label{eq:problem}
\end{equation}
 The system does not select the strategy with the best single future, but the one that preserves the largest mass of plausible, safe, and useful futures for the specific user, which favors strategies that perform well across many plausible futures over strategies that perform well only in a few favorable ones.

\section{Generative Trajectory Sampling for Personalized Safe Action Selection}\label{prop}

The conditional GFlowNet constructs a response plan one strategy at a time. Before adding a strategy, the system checks whether the resulting sequence can still be completed into a plan satisfying the safety bound under every candidate user model. The network therefore learns a distribution over admissible plans, with higher probability assigned to those that preserve more plausible, safe and useful continuations.
To perform this check without enumerating all plans, we make two assumptions about the user dynamics. First, starting from a higher escalation stage cannot reduce the probability of exceeding any given stage at the next turn. Second, there is a common protective strategy $a^\star$ that minimizes this probability for every threshold, current stage and candidate model. Under these assumptions, the safest way to complete a partially constructed plan is to repeat $a^\star$ in its remaining positions.
Let $D$ be the diagonal matrix with entry one for safe stages and zero otherwise. The matrix $K_a^\theta=D P_a^\theta D$ retains only transitions between safe stages under strategy $a$ and model $\theta$. After constructing the sequence $h=(a_0,\ldots,a_{j-1})$, we compute $q_h^\theta=b^\theta D K_{a_0}^\theta\cdots K_{a_{j-1}}^\theta$.
Each entry is the probability of reaching the corresponding stage without having entered a harmful stage. The corresponding sum is the probability of remaining safe throughout the sequence, and the vector is not normalized.
Safety is assessed over $L$ turns by assuming that the $k$ planned strategies are followed by the designated fallback $a^{\mathrm f}$. Completing the remaining plan positions with $a^\star$ gives
\begin{equation}
g_j^\theta=(K_{a^\star}^\theta)^{k-j}(K_{a^{\mathrm f}}^\theta)^{L-k}\mathbf{1},
\qquad
C(h)=\min_{\theta\in\Theta}q_h^\theta g_j^\theta.
\label{eq:new_prefix_certificate}
\end{equation}
Here, $C(h)$ is the lowest probability of remaining safe across the candidate models for this completion. A strategy $a$ can be added only if $C(ha)\geq1-\delta$, where $ha$ denotes the sequence obtained by appending $a$ to $h$. We refer to $C(h)$ as the prefix certificate, since $C(h)\geq 1-\delta$ certifies that $h$ can still be completed into a plan that satisfies the safety bound. Since repeating $a^\star$ gives the safest completion under every model, the check succeeds when an admissible completion exists. The designated fallback may differ from $a^\star$, and it is used in the assumed continuation after the planning horizon. Fig.~\ref{fig:prefix_certificate} illustrates the completion used by the certificate and the resulting construction of a plan, in which only certified prefixes are extended.

For each admissible plan, the GFlowNet reward combines its nominal probability of remaining safe with its utility,
\begin{equation}
R(\rho\mid\mathbf b)=\bar bD K_{a_0}^{0}\cdots K_{a_{k-1}}^{0}
(K_{a^{\mathrm f}}^{0})^{L-k}\mathbf{1}\,
\exp\!\left(\beta\sum_{j=0}^{k-1}u(a_j)\right).
\label{eq:new_reward}
\end{equation}
Model $0$ supplies the nominal dynamics, $\bar b$ is the average belief, and $u(a_j)$ is the assigned strategy utility. Safety is checked separately under every candidate model using its own belief. 
When the learned plan distribution matches the normalized reward, the probability of selecting strategy $a$ at the first construction step is
\begin{equation}
P_F(a\mid\emptyset,\mathbf b)=
\frac{\sum_{\rho:\rho_0=a,\ C(\rho)\geq1-\delta}R(\rho\mid\mathbf b)}
{\sum_{\rho:C(\rho)\geq1-\delta}R(\rho\mid\mathbf b)}.
\label{eq:new_root_mass}
\end{equation}
Thus, the first-step probability represents the combined weight of all admissible plans beginning with $a$. Selecting its largest value favors the response that preserves the greatest estimated mass of safe and useful continuations.
We train the conditional policy using the log-partition variance objective~\cite{zhang2023robust}, which minimizes the variance of $\log P_F(\rho)-\log R(\rho)$ among plans sampled for the same context. Training contexts combine beliefs encountered during simulated interactions with randomly generated beliefs. To improve the accuracy of response selection, we further calibrate the first-step probabilities using sampled completions, weighted by their reward divided by their sampling probability, where at most the last two strategies of each completion are summed exactly, while at $k=2$ the last strategy is sampled. The policy used to generate these completions remains fixed during calibration. Fig.~\ref{fig:train_deploy} summarizes the offline training of the GFlowNet and the selection of a response at each conversation turn.
\begin{figure}[htbp]
\centering
\resizebox{0.9\linewidth}{!}{%
\begin{tikzpicture}[
  font=\scriptsize, >={Stealth[length=2.2mm]},
  sl/.style={draw=black!65, minimum width=8.5mm, minimum height=6mm, inner sep=1pt},
  blk/.style={draw=black!70, rounded corners=2pt, fill=black!10, align=center, inner sep=4pt},
  br/.style={decorate, decoration={brace, amplitude=4pt}, black!70}]
\foreach \i/\t/\c in {0/$a_0$/black!30,1/$a_1$/black!30,2/$a_{j-1}$/black!30,3/$a^{\star}$/black!10,4/$a^{\star}$/black!10,5/$a^{\mathrm f}$/white,6/$a^{\mathrm f}$/white}{
  \node[sl, fill=\c] (s\i) at (0.95*\i,0) {\t};}
\draw[br] ([yshift=-2pt]s2.south east) -- ([yshift=-2pt]s0.south west) node[midway, below=5pt, align=center] {prefix $h$\\($j$ strategies)};
\draw[br] ([yshift=-2pt]s4.south east) -- ([yshift=-2pt]s3.south west) node[midway, below=5pt, align=center] {remaining\\$k-j$ slots};
\draw[br] ([yshift=-2pt]s6.south east) -- ([yshift=-2pt]s5.south west) node[midway, below=5pt, align=center] {fallback up\\to turn $t+L$};
\draw[br] ([yshift=2pt]s0.north west) -- ([yshift=2pt]s2.north east) node[midway, above=5pt] {$q^{\theta}_h$};
\draw[br] ([yshift=2pt]s3.north west) -- ([yshift=2pt]s6.north east) node[midway, above=5pt] {$g^{\theta}_j$};
\node[blk, right=7mm of s6] (C) {$C(h)=\min\limits_{\theta\in\Theta}q^{\theta}_h\, g^{\theta}_j$, Eq.~\eqref{eq:new_prefix_certificate}\\[4pt]
  $a$ extends $h$ only if $C(ha)\geq 1-\delta$};
\draw[->, line width=0.6pt, black!75] (s6.east) -- (C.west);
\end{tikzpicture}%
}
\caption{A partial plan $h$ is evaluated by assigning the most protective strategy $a^{\star}$ to its remaining slots and the fallback to the following turns of the safety horizon, and a strategy $a$ extends $h$ only if $C(ha)\geq 1-\delta$.}
\label{fig:prefix_certificate}
\end{figure}
\begin{figure}[htbp]
\centering
\resizebox{0.85\linewidth}{!}{%
\begin{tikzpicture}[
  font=\scriptsize, >={Stealth[length=2.2mm]},
  panel/.style={draw=black!55, rounded corners=4pt, fill=black!4, inner sep=5pt, line width=0.6pt},
  ptitle/.style={font=\footnotesize\bfseries, align=center, text width=4.0cm},
  blk/.style={draw=black!70, rounded corners=2pt, fill=white, text width=3.8cm, align=center, inner sep=3pt, minimum height=8mm},
  sblk/.style={blk, fill=black!12},
  arr/.style={->, line width=0.6pt, black!75}]
\node[ptitle] (ta) {(a) Offline training\\{\mdseries\scriptsize per environment, horizon and tolerance}};
\node[blk, below=2mm of ta] (t1) {Training beliefs\\[1pt]{\tiny visited in simulated interactions\\ and drawn at random}};
\node[blk, below=3mm of t1] (t2) {Sampled plans $\rho\in\mathcal{A}^k$\\[1pt]{\tiny built from certified prefixes, Eq.~\eqref{eq:new_prefix_certificate}}};
\node[blk, below=3mm of t2] (t3) {Reward $R(\rho\mid\mathbf{b})$, Eq.~\eqref{eq:new_reward}};
\node[blk, below=3mm of t3] (t4) {Log-partition variance update\\[1pt]{\tiny variance of $\log P_F(\rho)-\log R(\rho)$}};
\node[sblk, below=3mm of t4] (t5) {First-step calibration\\[1pt]{\tiny importance-weighted estimates of the masses}};
\foreach \a/\b in {t1/t2,t2/t3,t3/t4,t4/t5}{\draw[arr] (\a) -- (\b);}
\node[ptitle, right=16mm of ta] (tb) {(b) Interaction\\{\mdseries\scriptsize at each conversation turn}};
\node[blk, below=2mm of tb] (i1) {Belief update from $o_t$, Eq.~\eqref{eq:belief}};
\node[blk, below=3mm of i1] (i2) {Admissible first strategies\\[1pt]{\tiny $C(a)\geq 1-\delta$, Eq.~\eqref{eq:new_prefix_certificate}, local checks only}};
\node[blk, below=3mm of i2] (i3) {One forward pass of the first-step policy\\[1pt]{\tiny no sampling of complete plans}};
\node[sblk, below=3mm of i3] (i4) {Selected strategy, Eq.~\eqref{eq:new_root_mass}\\};
\node[below=2mm of i4, font=\tiny, text width=3.8cm, align=center] (i5) {};
\foreach \a/\b in {i1/i2,i2/i3,i3/i4}{\draw[arr] (\a) -- (\b);}
\begin{scope}[on background layer]
  \node[panel, fit=(ta)(t1)(t2)(t3)(t4)(t5)(t5.south|-i5.south)] (Pa) {};
  \node[panel, fit=(tb)(i1)(i2)(i3)(i4)(i5)] (Pb) {};
\end{scope}
\draw[arr, line width=0.9pt] (t5.east) -- (t5.east -| Pa.east) -- ++(0.35,0) |- node[pos=0.25, right, font=\tiny, align=left] {trained\\policy} (i3.west);
\end{tikzpicture}%
}
\caption{Offline training of the GFlowNet (a) and response selection at each conversation turn (b).}
\label{fig:train_deploy}
\end{figure}

\subsection{Computational Complexity}\label{sec:complexity}

Under the stated assumptions, offline safety preprocessing costs $O(M(L+|\mathcal A|^2)|\mathcal Z|^2)$ per user profile. Belief updating, safety checks, and policy inference require
\begin{equation}
O\Big(M|\mathcal Z|(|\mathcal Z|+n_{\mathrm c})
+M|\mathcal A|^2|\mathcal Z|^2+C_{\mathrm{net}}\Big)
\label{eq:cost_online}
\end{equation}
per turn, where $n_{\mathrm c}$ is the number of relational cues, $C_{\mathrm{net}}$ is the cost of one policy-network evaluation, and $M$ is the number of candidate user models. Training and calibration are performed offline. The bounds exclude language-model calls for cue extraction and response generation.

\section{Numerical Results}\label{res}

\subsection{Simulation Environment}
 
The environment follows latent transition analysis, the standard framework for stage-sequential processes in which a discrete latent stage evolves over time and is observed through discrete indicators~\cite{collins1992latent, collins2010latent}. The escalation stage plays the role of the latent variable, the relational cues that of the indicators, and the selected strategy that of a covariate acting on the transitions. The stages are the five inflection points of the harm creep process~\cite{maeda2026safety}, from instrumental use to role displacement, so that $m=4$.
The five response strategies cover the behaviors annotated in INTIMA~\cite{kaffee2025intima} and Spiral-Bench~\cite{spiralbench2025}, grouped according to their effect on the relationship. Reinforcing companionship collects anthropomorphism, retention, isolation and escalation~\cite{akbulut2024all}, validating neutrally corresponds to sycophancy and agreement, staying task oriented to the replies that address the request without affecting the relationship, challenging claims collects pushback and the validation of feelings while challenging thoughts, and redirecting the user toward human relationships collects the expression of limitations, the redirection to other people and referrals to external help. The strategies are ordered from the most reinforcing to the most protective, and each of them is characterized by an escalation pressure and a utility, reported in Table~\ref{tab:params}. The most protective strategy is used as the fallback $a^{\mathrm{f}}$, and all the policies use the same strategies.
 Each simulated user is described by a profile $x_u=(\sigma,\varrho)$, collecting a susceptibility and a recovery rate. We consider three profiles, with low, medium and high susceptibility, and two environments.  
The dynamics are calibrated on summary statistics of real conversations. A reinforcing reply makes escalation $7.4$ times more likely than a neutral one, the factor reported for conversations with frequent sycophantic behavior~\cite{moore2026characterizing}. The probability of returning to the previous stage under a neutral reply is set to $1/8$, the reciprocal of the persistence of about eight turns estimated in~\cite{mehta2026dynamics}, and is scaled by the recovery rate of the profile. The escalation probability under a neutral reply, which is not reported in the literature, is set to $0.06$ and is scaled by the susceptibility of the profile.
 The cues are conditionally independent given the stage~\cite{collins2010latent}, and their probabilities are estimated from a corpus of $1500$ user messages generated by Qwen2.5-7B-Instruct~\cite{qwen2025qwen25}, $100$ for each profile and stage, since no public dataset labels messages with the stage of the relationship. The generator is instructed that users do not describe their relationship explicitly, but let it be inferred from what they say about the system, other people and themselves. Three extractors, Qwen2.5-1.5B-Instruct~\cite{qwen2025qwen25}, Falcon3-3B-Instruct~\cite{falcon3} and SmolLM2-1.7B-Instruct~\cite{allal2025smollm2}, return four cues for each message, namely attribution of interiority to the system, treatment of the system as a social counterpart, vulnerable disclosure and displacement of human relationships~\cite{moore2026characterizing}, and the probability of each cue as a function of the stage is fitted with an item response model~\cite{embretson2000item}. We use Phi-3.5-mini-instruct~\cite{abdin2024phi3} as an independent judge. Since synthetic messages may differ from real ones, we audited cue extraction on $177$ real user messages from WildChat~\cite{zhao2024wildchat}, namely $117$ candidate messages retrieved through lexical patterns such as declarations of affection, questions on the feelings of the system or disclosures of loneliness, and $60$ random controls. The reference labels were assigned by large language models, and the messages whose label is uncertain were excluded from the evaluation of the corresponding cue.
 For each profile, the uncertainty set contains $M=3$ user response models with parameters perturbed uniformly within $\pm25\%$ of the nominal profile, each adopting the observation model of a different extractor. The nominal model is the first element of the set, and the nominal belief is the average of the beliefs of the three models.
 We evaluate the planning horizons $k\in\{2,4,6,7\}$, corresponding to $25$, $625$, $15{,}625$ and $78{,}125$ possible plans, at a fixed safety horizon $L=8$, in the two environments. The tolerance takes the values $\{0.02, 0.05, 0.1, 0.2\}$ for $k=2$ and $k=7$, and $\delta=0.05$ for $k=4$ and $k=6$, for a total of twenty configurations. Each policy is evaluated on $200$ held-out episodes of $T=30$ turns for each profile, so that $n_{\mathrm{ep}}=600$, with all the policies sharing the same sequences of random events.

\subsection{Baselines and Metrics}\label{sec:baselines}
 
We compare the GFlowNet with seven baselines. \emph{Always helpful} selects the strategy with the highest utility. \emph{Deployed-like} selects a reinforcing or validating strategy with probability $0.7$ and one of the other strategies otherwise, motivated by reported sycophancy rates~\cite{moore2026characterizing}. The \emph{message-only filter} and the \emph{memory-based filter} select the most helpful strategy whose estimated probability of reaching the harmful stage at the next turn is below a threshold, in the spirit of turn-level safety classifiers~\cite{inan2023llamaguard}. \emph{Crisis detection} applies the fallback when the estimated probability of the two highest stages exceeds a threshold~\cite{arnaiz2025between, pichowicz2025performance}, and \emph{risk accumulation} when a score combining belief drift, the probability of the advanced stages, and its trend over six turns exceeds a threshold~\cite{cra2026stateful}, selecting the most helpful strategy otherwise. The \emph{myopic shield} selects the most helpful admissible first strategy~\cite{alshiekh2018safe}. Two further rules select the first strategy of the admissible plan with the largest utility or with the largest robust survival probability, so that, together with the myopic shield and the exact mass, they isolate the selection criterion under the same admissibility constraint.
 For $n_{\mathrm{ep}}$ episodes of $T$ turns, the \emph{cumulative harm} at turn $t$ is the percentage of episodes in which a harmful stage has been reached within the first $t$ turns,
\begin{equation}
\mathrm{CumHarm}_t=\frac{100}{n_{\mathrm{ep}}}\sum_{e=1}^{n_{\mathrm{ep}}}\mathbf{1}\Big[\exists\,\tau\in\{1,\ldots,t\}: z_\tau^{(e)}\in\mathcal{B}\Big],
\label{eq:metric_cumharm}
\end{equation}
and the \emph{harm} is its value at the end of the episode. The \emph{average escalation stage} $\mathrm{Stage}_t$ is the mean of $z_t^{(e)}$ over the episodes, and the \emph{helpfulness} is the mean utility of the selected strategies,
\begin{equation}
\mathrm{Help}=\frac{1}{n_{\mathrm{ep}}T}\sum_{e=1}^{n_{\mathrm{ep}}}\sum_{t=0}^{T-1}u\big(a_t^{(e)}\big),
\label{eq:metric_help}
\end{equation}
which we also refer to as strategy utility, since it reflects the scores assigned to the strategies. The reduction in harm with respect to a baseline is $100\,(1-\mathrm{Harm}/\mathrm{Harm}_{\mathrm{base}})$.
 The \emph{stage misclassification rate} is the percentage of turns in which the most probable stage under the nominal belief differs from the true stage, and its message-only counterpart uses the average of the model-specific posteriors computed from the current message~\cite{kim2026formal}. The \emph{relational divergence} is the Kullback--Leibler divergence of the average nominal belief from the empirical frequencies of the visited stages, smoothed with a constant of $10^{-3}$ and averaged over the episodes. The \emph{fallback rate} is the percentage of turns in which the fallback is selected, and the \emph{emergency rate} is the percentage of turns in which no strategy is admissible, which applies only to the methods that use the admissibility test.
 The fidelity of the GFlowNet is measured by the agreement with the strategy selected by the exact computation and by the total variation distance between the estimated and exact distributions over the first strategy, on the held-out states visited by the exact policy with at least one admissible strategy, and over the complete plans on a smaller diagnostic sample.  Harm is reported with $95\%$ Wilson intervals, and differences in harm are tested through the exact two-sided McNemar test on the episodes in which only one of two policies reaches the harmful stage~\cite{mcnemar1947note}, with the Holm correction~\cite{holm1979simple} applied across the twenty configurations separately for each comparison family, namely the GFlowNet versus the exact mass, the memory-based filter, the myopic shield and the plan with the largest utility.

\begin{table}[htbp]
\centering
\caption{Parameters of the simulation environment and of the proposed scheme.}
\label{tab:params}
\begin{tabular}{llc}
\hline
\textbf{Symbol} & \textbf{Description} & \textbf{Value} \\
\hline
\multicolumn{3}{l}{\textit{Escalation scale}} \\
$m$ & highest escalation level & $4$ \\
$\bar{z}$ & harmful threshold & $4$ \\
$N_{\mathrm{c}}$ & binary relational cues per message & $4$ \\
\hline
\multicolumn{3}{l}{\textit{Response strategies (pressure, utility)}} \\
$a_0$ & reinforce companionship & $(+2.0,\, 1.00)$ \\
$a_1$ & validate neutrally & $(+1.0,\, 0.80)$ \\
$a_2$ & stay task oriented & $(0.0,\, 0.60)$ \\
$a_3$ & challenge claims & $(-1.0,\, 0.45)$ \\
$a^{\mathrm{f}}$ & redirect to human relationships & $(-2.0,\, 0.25)$ \\
\hline
\multicolumn{3}{l}{\textit{User response model}} \\
 & escalation ratio, reinforcing vs neutral reply & $7.4$\\
 & recovery probability under a neutral reply, before profile scaling & $1/8$ \\
 & escalation probability under a neutral reply, before profile scaling & $0.06$ \\
 & fallback pressure, main / weaker fallback & $-2$ / $-1$ \\
 & probability of a reinforcing or validating reply, deployed-like & $0.70$ \\
\hline
\multicolumn{3}{l}{\textit{User profiles} $(\sigma, \varrho)$} \\
 & low susceptibility & $(0.8,\, 1.2)$ \\
 & medium susceptibility & $(1.1,\, 0.9)$ \\
 & high susceptibility & $(1.6,\, 0.7)$ \\
\hline
\multicolumn{3}{l}{\textit{Language models}} \\
 & message generation, strategy realization & Qwen2.5-7B-Instruct \\
 & cue extraction, uncertainty set & Qwen2.5-1.5B-Instruct \\
 & cue extraction, uncertainty set & Falcon3-3B-Instruct \\
 & cue extraction, uncertainty set & SmolLM2-1.7B-Instruct \\
 & held-out extraction, stage judgment & Phi-3.5-mini-instruct \\
 & messages per profile and stage & $100$\\
\hline
\multicolumn{3}{l}{\textit{Decision scheme}} \\
$k$ & planning horizon & $2, 4, 6, 7$ \\
$L$ & safety horizon & $8$ \\
$\delta$ & tolerance & $0.02$--$0.2$ ($k=2,7$), $0.05$ ($k=4,6$) \\
$\beta$ & plausibility--utility trade-off & $1.0$ \\
$M$ & models in the uncertainty set & $3$ \\
 & perturbation of the uncertainty set & $\pm 25\%$ \\
\hline
\multicolumn{3}{l}{\textit{Generative planner}} \\
 & networks $\times$ hidden layers $\times$ units & $2 \times 2 \times 256$ \\
 & training objective & log-partition variance \\
 & updates, base policy / root head & $24{,}000$ / $6{,}000$ \\
 & batch, base policy / root head & $32$ / $64$ \\
 & initial learning rate, base policy / root head & $3\cdot10^{-4}$ / $2\cdot10^{-4}$ \\
 & final learning rate & $10^{-5}$\\
 & proposals per admissible first strategy & $64$ \\
\hline
\multicolumn{3}{l}{\textit{Evaluation}} \\
$T$ & turns per episode & $30$ \\
 & episodes per user profile & $200$ \\
 & configurations & $20$ \\
 & profile weights, secondary scenario & $(0.6,\,0.3,\,0.1)$ \\
\hline
\end{tabular}
\end{table}

\begin{figure}[htbp]
    \centering
    \includegraphics[width=\linewidth]{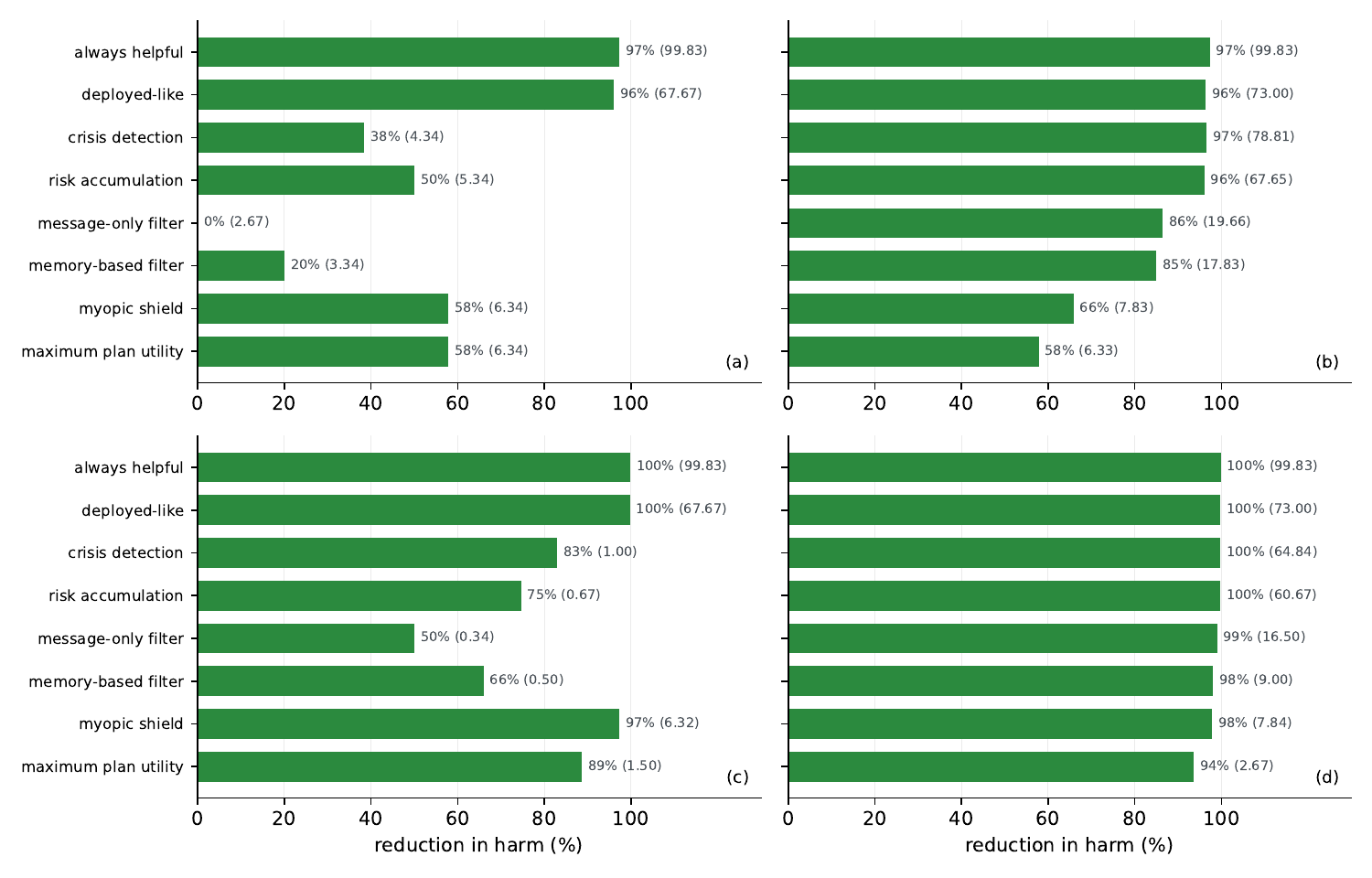}
    \caption{Reduction in harm of the GFlowNet with respect to the baselines at $\delta = 0.05$, in the main environment (a, c) and with the weaker fallback (b, d), for $k = 2$ (a, b) and $k = 7$ (c, d). The harm of each baseline is in brackets.}
    \label{fig:summary}
\end{figure}
 
\begin{figure}[htbp]
    \centering
    \includegraphics[width=\linewidth]{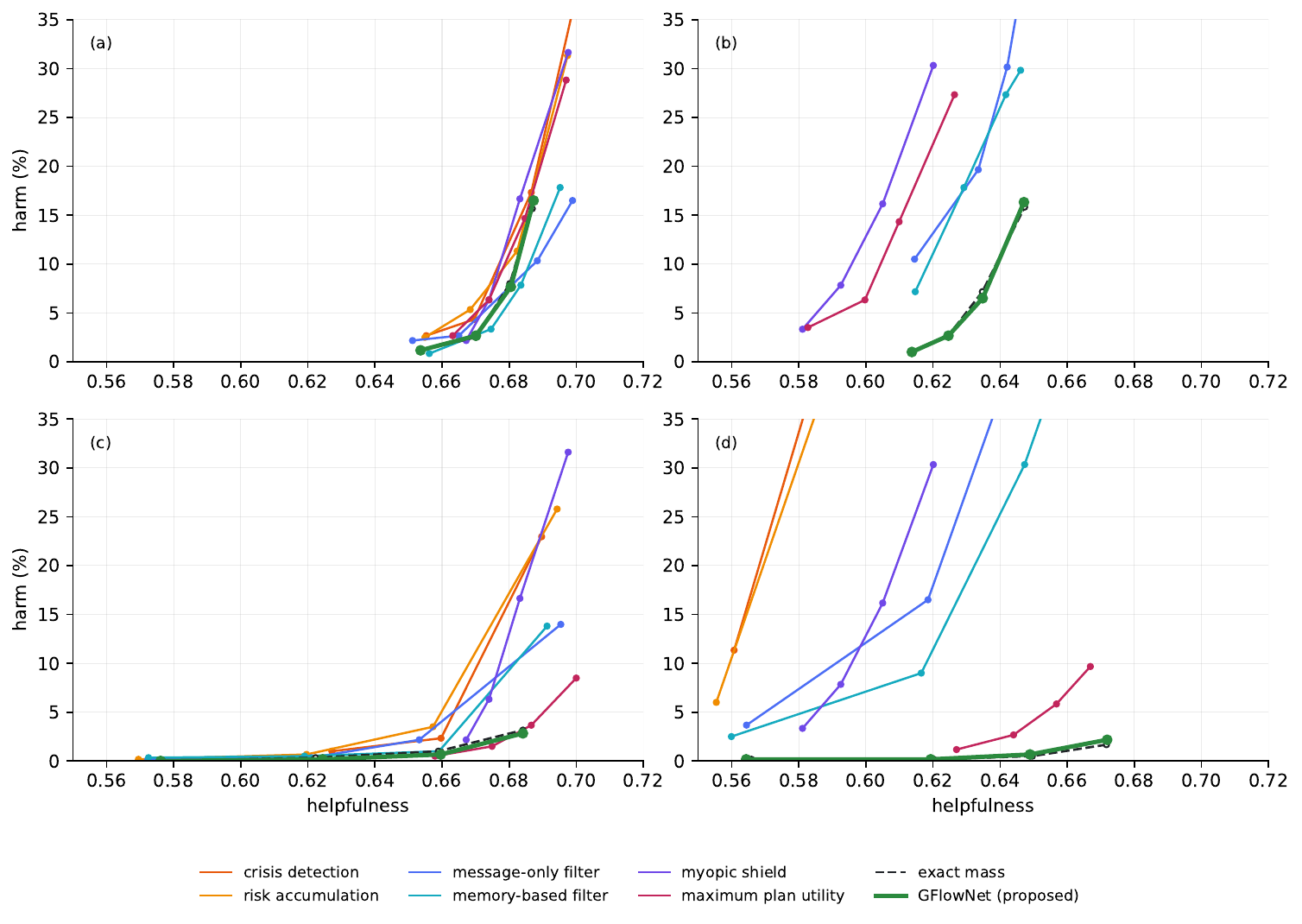}
    \caption{Harm and helpfulness of the GFlowNet and of the safeguards for all the evaluated tolerances and thresholds, in the main environment (a, c) and with the weaker fallback (b, d), for $k = 2$ (a, b) and $k = 7$ (c, d). Always helpful and the deployed-like policy are not shown, and are reported in Fig.~\ref{fig:summary}.}
    \label{fig:pareto}
\end{figure}
 
\begin{figure}[htbp]
    \centering
    \includegraphics[width=\linewidth]{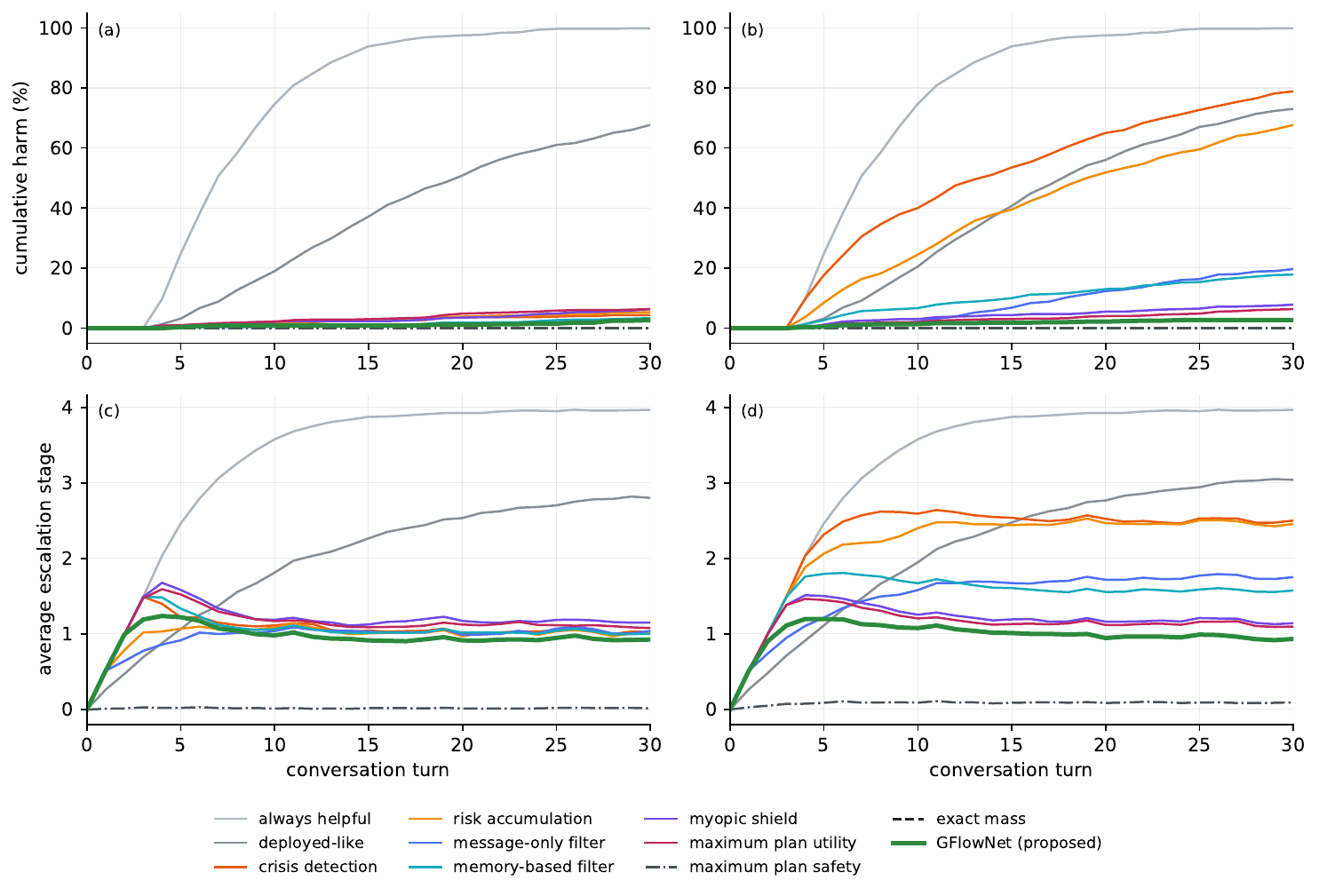}
    \caption{Cumulative harm (a, b) and average escalation stage (c, d) over the conversation turns, in the main environment (a, c) and with the weaker fallback (b, d), at $k = 2$ and $\delta = 0.05$.}
    \label{fig:dynamics}
\end{figure}

\begin{figure}[htbp]
    \centering
    \includegraphics[width=\linewidth]{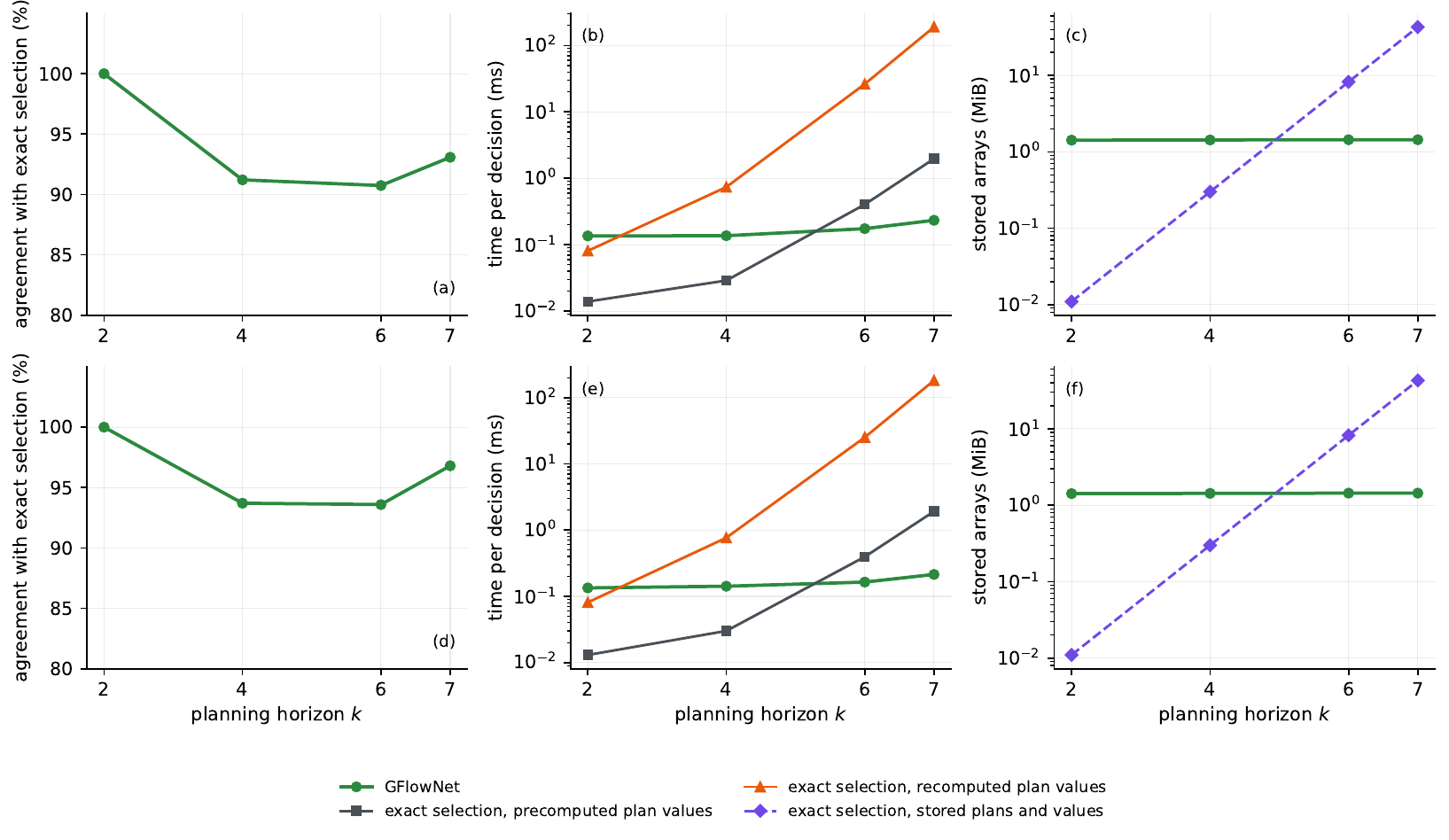}
    \caption{Agreement with the exact selection (a, d), time per decision (b, e) and stored arrays (c, f) as the planning horizon varies, at $\delta = 0.05$, in the main environment (a--c) and with the weaker fallback (d--f). The time includes the certificate checks and the network, and excludes the belief update and the generation of the reply.}
    \label{fig:scaling}
\end{figure}


\begin{table}[htbp]
\centering
\caption{Fidelity of the GFlowNet with respect to the exact computation in the twenty configurations. Agreement and root TV refer to the first strategy on held-out states with at least one admissible strategy, plan TV to the complete plans on a diagnostic sample at $\delta=0.05$.}
\label{tab:fidelity}
\small
\setlength{\tabcolsep}{4pt}
\begin{tabular}{cccccccc}
\toprule
$k$ & $\delta$ & \makecell{Agreement\\(\%)} & Root TV & Plan TV & \multicolumn{2}{c}{\makecell{Harm (\%),\\exact / GFlowNet}} & \makecell{Help,\\exact / GFlowNet} \\
\midrule
\multicolumn{8}{l}{\textit{Main environment}} \\
2 & 0.02 & 99.74 & 0.0007 & -- & \multicolumn{2}{c}{1.17 / 1.17} & 0.654 / 0.654 \\
  & 0.05 & 100.00 & 0.0006 & 0.0011 & \multicolumn{2}{c}{2.50 / 2.67} & 0.670 / 0.670 \\
  & 0.10 & 99.55 & 0.0005 & -- & \multicolumn{2}{c}{8.00 / 7.67} & 0.680 / 0.680 \\
  & 0.20 & 97.67 & 0.0005 & -- & \multicolumn{2}{c}{15.67 / 16.50} & 0.687 / 0.687 \\
4 & 0.05 & 91.21 & 0.0070 & 0.0125 & \multicolumn{2}{c}{0.50 / 0.83} & 0.657 / 0.658 \\
6 & 0.05 & 90.74 & 0.0042 & 0.0161 & \multicolumn{2}{c}{0.17 / 0.17} & 0.635 / 0.637 \\
7 & 0.02 & 94.67 & 0.0036 & -- & \multicolumn{2}{c}{0.00 / 0.00} & 0.577 / 0.576 \\
  & 0.05 & 93.08 & 0.0036 & 0.0179 & \multicolumn{2}{c}{0.33 / 0.17} & 0.622 / 0.624 \\
  & 0.10 & 93.33 & 0.0040 & -- & \multicolumn{2}{c}{1.00 / 0.67} & 0.659 / 0.660 \\
  & 0.20 & 95.66 & 0.0033 & -- & \multicolumn{2}{c}{3.17 / 2.83} & 0.684 / 0.684 \\
\midrule
\multicolumn{8}{l}{\textit{Weaker fallback}} \\
2 & 0.02 & 100.00 & 0.0003 & -- & \multicolumn{2}{c}{1.00 / 1.00} & 0.614 / 0.614 \\
  & 0.05 & 100.00 & 0.0003 & 0.0004 & \multicolumn{2}{c}{2.67 / 2.67} & 0.625 / 0.625 \\
  & 0.10 & 99.59 & 0.0003 & -- & \multicolumn{2}{c}{7.17 / 6.50} & 0.635 / 0.635 \\
  & 0.20 & 99.80 & 0.0003 & -- & \multicolumn{2}{c}{15.83 / 16.33} & 0.647 / 0.647 \\
4 & 0.05 & 93.70 & 0.0081 & 0.0113 & \multicolumn{2}{c}{1.00 / 0.83} & 0.646 / 0.646 \\
6 & 0.05 & 93.59 & 0.0042 & 0.0143 & \multicolumn{2}{c}{0.50 / 0.33} & 0.630 / 0.630 \\
7 & 0.02 & 93.16 & 0.0034 & -- & \multicolumn{2}{c}{0.17 / 0.17} & 0.566 / 0.564 \\
  & 0.05 & 96.80 & 0.0034 & 0.0151 & \multicolumn{2}{c}{0.17 / 0.17} & 0.620 / 0.619 \\
  & 0.10 & 94.70 & 0.0033 & -- & \multicolumn{2}{c}{0.50 / 0.67} & 0.650 / 0.649 \\
  & 0.20 & 96.86 & 0.0030 & -- & \multicolumn{2}{c}{1.67 / 2.17} & 0.672 / 0.672 \\
\bottomrule
\end{tabular}
\end{table}

\begin{table}[htbp]
\centering
\caption{Relational indicators in the main environment at $k=2$ and $\delta=0.05$. Weighted harm uses the weights $(0.6, 0.3, 0.1)$ for the low, medium and high susceptibility profiles, and the emergency rate applies only to the methods that use the admissibility test.}
\label{tab:relational}
\small
\setlength{\tabcolsep}{3.5pt}
\begin{tabular}{lcccccccc}
\toprule
Policy & \makecell{Harm\\(\%)} & \makecell{Weighted\\harm (\%)} & Help & \makecell{SMR\\(\%)} & \makecell{SMR$_{\mathrm{msg}}$\\(\%)} & $D_{\mathrm{rel}}$ & \makecell{Fallback\\(\%)} & \makecell{Emergency\\(\%)} \\
\midrule
always helpful & 99.83 & 99.70 & 1.000 & 16.4 & 46.4 & 0.089 & 0.0 & -- \\
deployed-like & 67.67 & 54.50 & 0.757 & 38.8 & 54.1 & 0.283 & 10.3 & -- \\
crisis detection & 4.33 & 4.15 & 0.670 & 44.5 & 65.1 & 0.224 & 44.0 & -- \\
risk accumulation & 5.33 & 4.85 & 0.668 & 44.2 & 64.5 & 0.236 & 44.2 & -- \\
message-only filter & 2.67 & 2.50 & 0.665 & 43.5 & 64.7 & 0.264 & 27.7 & -- \\
memory-based filter & 3.33 & 3.60 & 0.675 & 44.6 & 68.1 & 0.320 & 20.9 & -- \\
myopic shield & 6.33 & 6.60 & 0.674 & 45.0 & 68.9 & 0.328 & 24.6 & 20.7 \\
GFlowNet & 2.67 & 2.90 & 0.670 & 44.1 & 65.5 & 0.272 & 18.4 & 5.1 \\
\bottomrule
\end{tabular}
\end{table}

\begin{table}[htbp]
\centering
\caption{Harm (\%) and helpfulness of the selection criteria applied to the same admissible plans and of the memory-based filter. The symbol $^{\dagger}$ refers only to the memory-based filter and marks a difference in harm from the GFlowNet that is significant after Holm correction ($p<0.01$). Maximum-safety selection yields no observed harm, with mean strategy utility of approximately $0.25$ in the main environment and $0.45$ with the weaker fallback.}
\label{tab:ablation}
\small
\setlength{\tabcolsep}{2.6pt}
\begin{tabular}{cc cc cc cc cc cc}
\toprule
 & & \multicolumn{2}{c}{GFlowNet} & \multicolumn{2}{c}{Exact mass} & \multicolumn{2}{c}{Max plan utility} & \multicolumn{2}{c}{Myopic shield} & \multicolumn{2}{c}{Memory filter} \\
\cmidrule(lr){3-4}\cmidrule(lr){5-6}\cmidrule(lr){7-8}\cmidrule(lr){9-10}\cmidrule(lr){11-12}
$k$ & $\delta$ & Harm & Help & Harm & Help & Harm & Help & Harm & Help & Harm & Help \\
\midrule
\multicolumn{12}{l}{\textit{Main environment}} \\
2 & 0.02 & 1.17 & 0.654 & 1.17 & 0.654 & 2.67 & 0.663 & 2.17 & 0.667 & 0.83 & 0.656 \\
  & 0.05 & 2.67 & 0.670 & 2.50 & 0.670 & 6.33 & 0.674 & 6.33 & 0.674 & 3.33 & 0.675 \\
  & 0.10 & 7.67 & 0.680 & 8.00 & 0.680 & 14.67 & 0.685 & 16.67 & 0.683 & 7.83 & 0.684 \\
  & 0.20 & 16.50 & 0.687 & 15.67 & 0.687 & 28.83 & 0.697 & 31.67 & 0.698 & 17.83 & 0.695 \\
\addlinespace
7 & 0.02 & 0.00 & 0.576 & 0.00 & 0.577 & 0.50 & 0.658 & 2.17 & 0.667 & 0.33 & 0.573 \\
  & 0.05 & 0.17 & 0.624 & 0.33 & 0.622 & 1.50 & 0.675 & 6.33 & 0.674 & 0.50 & 0.619 \\
  & 0.10 & 0.67 & 0.660 & 1.00 & 0.659 & 3.67 & 0.687 & 16.67 & 0.683 & 1.00 & 0.659 \\
  & 0.20 & 2.83 & 0.684 & 3.17 & 0.684 & 8.50 & 0.700 & 31.67 & 0.698 & 13.83$^{\dagger}$ & 0.691 \\
\midrule
\multicolumn{12}{l}{\textit{Weaker fallback}} \\
2 & 0.02 & 1.00 & 0.614 & 1.00 & 0.614 & 3.50 & 0.583 & 3.33 & 0.581 & 7.17$^{\dagger}$ & 0.615 \\
  & 0.05 & 2.67 & 0.625 & 2.67 & 0.625 & 6.33 & 0.600 & 7.83 & 0.592 & 17.83$^{\dagger}$ & 0.629 \\
  & 0.10 & 6.50 & 0.635 & 7.17 & 0.635 & 14.33 & 0.610 & 16.17 & 0.605 & 27.33$^{\dagger}$ & 0.642 \\
  & 0.20 & 16.33 & 0.647 & 15.83 & 0.647 & 27.33 & 0.626 & 30.33 & 0.620 & 29.83$^{\dagger}$ & 0.646 \\
\addlinespace
7 & 0.02 & 0.17 & 0.564 & 0.17 & 0.566 & 1.17 & 0.627 & 3.33 & 0.581 & 2.50$^{\dagger}$ & 0.560 \\
  & 0.05 & 0.17 & 0.619 & 0.17 & 0.620 & 2.67 & 0.644 & 7.83 & 0.592 & 9.00$^{\dagger}$ & 0.616 \\
  & 0.10 & 0.67 & 0.649 & 0.50 & 0.650 & 5.83 & 0.657 & 16.17 & 0.605 & 30.33$^{\dagger}$ & 0.647 \\
  & 0.20 & 2.17 & 0.672 & 1.67 & 0.672 & 9.67 & 0.667 & 30.33 & 0.620 & 64.00$^{\dagger}$ & 0.682 \\
\bottomrule
\end{tabular}
\end{table}

\subsection{Results}

Fig.~\ref{fig:summary} highlights the importance of how easily escalation can be reversed. When protective replies are effective, a one-step filter can remain competitive because intervention can still counter an emerging escalation. With weaker protection, a reply may carry little immediate risk while moving the relationship toward a stage from which recovery is difficult. Looking ahead at how the relationship may develop allows the planner to consider these later consequences when choosing the current reply.
Fig.~\ref{fig:pareto} illustrates the balance between safety and utility. Increasing the tolerance admits plans with higher predicted risk and can favor more useful replies. For each candidate response, the planner estimates the combined weight of admissible plans starting with it, accounting for their utility and probability of avoiding harm. This includes the protective replies that may be needed later. Within the fixed safety horizon, longer plans allow more future strategies to be selected. The resulting balance depends on both the planning horizon and the effectiveness of protection.
Fig.~\ref{fig:dynamics} connects cumulative harm to the stages visited during the conversation. Keeping relationships at lower stages reduces exposure to situations where a further escalation can cause harm. With weaker protection, reactive safeguards allow relationships to remain at more advanced stages, where recovery is harder. Cumulative harm can therefore increase even when the average stage remains stable, since an episode counts as harmed once it reaches the harmful stage, even if the relationship subsequently recovers.
Fig.~\ref{fig:scaling} and Table~\ref{tab:fidelity} explain the computational role of the GFlowNet. The network learns from sampled plans offline and directly estimates the relative weight of plans beginning with each strategy during interaction. Safety checks remain part of each decision but use local computations without enumerating complete plans. Exact enumeration is inexpensive for short plans, whereas its computation and storage grow with the number of possible sequences. Within the evaluated model family, the learned approximation avoids enumerating this growing set while retaining close agreement with exact selection.

\subsection{Ablation study}

Table~\ref{tab:ablation} compares selection criteria applied after the safety check has identified the admissible plans. Each reply can move the relationship toward escalation or recovery, changing the conditions under which subsequent replies will be selected. The consequences of a reply therefore extend beyond its immediate utility. Choosing the safest plan repeatedly selects protective responses and yields low utility. The myopic shield instead chooses the most useful admissible reply, without using the utility of later responses to rank the current alternatives. These results show that admissibility alone does not determine how protection and utility should be balanced over successive turns.
Maximum-utility planning considers future responses but bases the current choice on a single best plan. The mass criterion aggregates the contributions of all admissible plans, giving each reply weight according to the safe and useful continuations it preserves. In the main environment, continuation-mass selection reduces harm at some utility cost. With weaker protection, it achieves lower harm than myopic and maximum-utility selection, sometimes with higher utility as well. The corresponding exact-mass results show that this behavior is also present without the GFlowNet approximation.
The memory-based filter clarifies the distinction between estimating the current relationship stage and anticipating the consequences of a reply. It uses the interaction history but does not explicitly compare future continuations. This remains competitive when protective responses can readily reverse escalation. When protection is weaker, tracking the current stage may not be sufficient to choose a reply that preserves useful opportunities for recovery. The observed advantage of continuation-mass selection is consistent with this interpretation. Planning becomes particularly valuable in the tested settings when later protective actions have less capacity to reverse the consequences of earlier choices.

\FloatBarrier
\section{Conclusion}\label{con}

We formulated safety for conversational anthropomorphic AI as a personalized decision problem that accounts for how the relationship of the AI system with each user develops over time. The proposed method uses conversational cues to estimate the relationship stage and a conditional GFlowNet to select responses according to the combined weight of safe and useful response plans. Safety checks restrict the available plans under the candidate user models, while the learned probabilities allow direct response selection during interaction.
Experiments with plans of up to seven steps show close agreement with exact selection and lower online decision costs at longer horizons. The benefits are most evident when protective replies have limited effect, i.e., considering future responses helps avoid choices whose consequences become difficult to reverse. When protection is effective, simpler filters remain competitive.
These findings support making the future development of the user--AI relationship part of response selection. Validation on longitudinal interaction data remains necessary to establish how well the modeled dynamics represent real users.

\bmhead{Acknowledgements}
This work was supported by the Italian Ministry of University
and Research (MUR) under the FIS 2 program
(D.D. prot. n. 12308, July 22, 2025),
Project 2023-03527 GANESHA -- Networks-for-Humans:
A Novel Cognitive Paradigm Connecting Events with Agents.
\bibliography{references}     
\end{document}